\documentclass[
 reprint,
superscriptaddress,
amsmath,amssymb,
]{revtex4-2}
\usepackage{float} 
\usepackage{graphicx}
\usepackage{dcolumn}
\usepackage{bm}

\usepackage{array}
\usepackage{hyperref}  

\usepackage[T1]{fontenc}
\DeclareMathAlphabet{\mathpzc}{T1}{pzc}{m}{it}
\newcommand{\ts}{\textsuperscript}

\makeatletter
\def\blfootnote{\gdef\@thefnmark{}\@footnotetext}
\makeatother

\begin{document}

\preprint{}

\title{A model of phase-coupled delay equations for
the dynamics of word usage}

\author{Alejandro Pardo Pintos}
\affiliation{Universidad de Buenos Aires, Facultad de Ciencias Exactas y Naturales, Departamento de Física, Ciudad Universitaria, 1428 Buenos Aires, Argentina.}
\affiliation{CONICET - Universidad de Buenos Aires, Instituto de Física Interdisciplinaria y Aplicada (INFINA), Ciudad Universitaria, 1428 Buenos Aires, Argentina.}
\author{Diego E Shalom}
\affiliation{Universidad de Buenos Aires, Facultad de Ciencias Exactas y Naturales, Departamento de Física, Ciudad Universitaria, 1428 Buenos Aires, Argentina.}
\affiliation{CONICET - Universidad de Buenos Aires, Instituto de Física Interdisciplinaria y Aplicada (INFINA), Ciudad Universitaria, 1428 Buenos Aires, Argentina.}
\affiliation{Neuroscience Lab, Universiy Torcuato Di Tella, Av. Pres. Figueroa Alcorta 7350, C1428 CABA, Argentina.}
\author{Enzo Tagliazucchi}
\affiliation{Universidad de Buenos Aires, Facultad de Ciencias Exactas y Naturales, Departamento de Física, Ciudad Universitaria, 1428 Buenos Aires, Argentina.}
\affiliation{CONICET - Universidad de Buenos Aires, Instituto de Física Interdisciplinaria y Aplicada (INFINA), Ciudad Universitaria, 1428 Buenos Aires, Argentina.}
\affiliation{Latin American Brain Health Institute (BrainLat), Universidad Adolfo Ibañez, Av. Diag. Las Torres 2640, 7941169 Santiago, Chile.}
\author{Gabriel Mindlin}
\affiliation{Universidad de Buenos Aires, Facultad de Ciencias Exactas y Naturales, Departamento de Física, Ciudad Universitaria, 1428 Buenos Aires, Argentina.}
\affiliation{CONICET - Universidad de Buenos Aires, Instituto de Física Interdisciplinaria y Aplicada (INFINA), Ciudad Universitaria, 1428 Buenos Aires, Argentina.}
\author{Marcos Trevisan}
\email{marcos@df.uba.ar}
\affiliation{Universidad de Buenos Aires, Facultad de Ciencias Exactas y Naturales, Departamento de Física, Ciudad Universitaria, 1428 Buenos Aires, Argentina.}
\affiliation{CONICET - Universidad de Buenos Aires, Instituto de Física Interdisciplinaria y Aplicada (INFINA), Ciudad Universitaria, 1428 Buenos Aires, Argentina.}

\date{\today}

\begin{abstract}

Word use presents regular oscillations mounted over slowly varying trends. These oscillations have been recently interpreted in terms of fashion-like cycles of interest and saturation, and modelled using a logistic equation with distributed delay. 
Here we show that the communities of semantically related words are partially synchronized. To account for this, we model the words of each community using logistic equations connected with a Kuramoto coupling. In this way, we test the simple hypothesis that the change in the occurrence of a word depends linearly on the occurrence of its semantic neighbours. We show that this simple model reproduces the coherence observed in the experimental communities using a single global coupling across multiple languages, regardless of the network topology. Our results build confidence on a universal model of language usage based on the interaction between  cognitive foces and the sociocultural context.

\end{abstract}


\blfootnote {Author contributions: GM and MT designed the research; APP and DES analyzed the data; APP, ET and MT wrote the paper.  }

\maketitle

\section*{Introduction}

\begin{figure*}
\includegraphics[width=15.5cm]{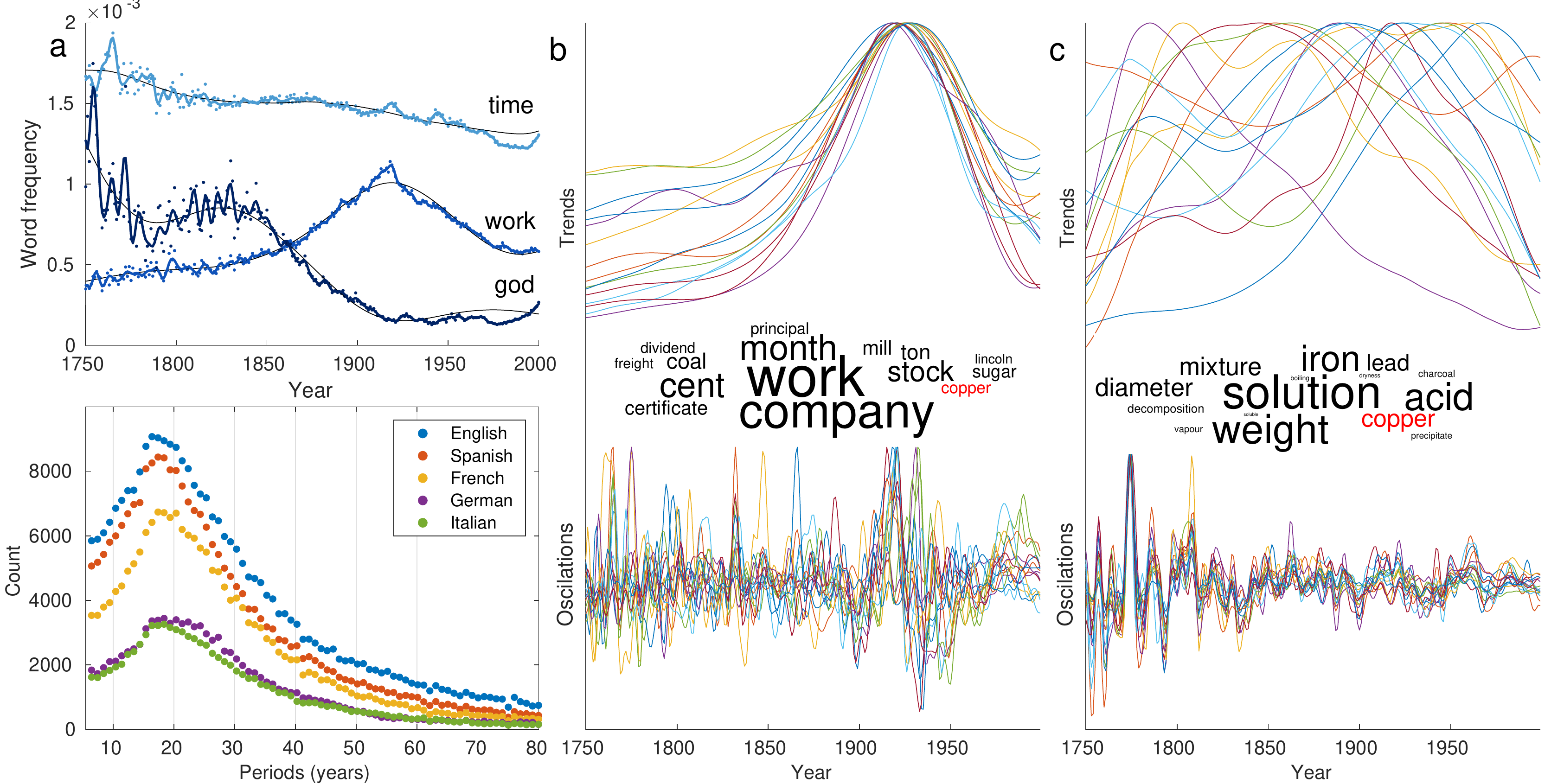}
\caption{ \textbf{Word usage consists on cycles mounted over slowly varying trends.}  (a) Frequency of usage of the words {\em time, work}, and {\em god} over the last three centuries, showing oscillations mounted on a slowly varying trend (black). Wavelet analysis revealed the dominance of oscillations with periods $\sim$ 15 years across languages (bottom panel). (b) We analyzed trends and oscillations separately: for each community of words, we show the normalized trends in the upper panel and the normalized oscillations in the lower panel. Similar trends correspond to keywords of sociocultural periods. The trends are similar, and the oscillations are more variable. (c) Similar oscillations correspond to semantically related words, as shown by the chemistry group. In this case, the oscillations of the words are synchronized, but the trends present high variability.}
\label{fig2}
\end{figure*}

Most language changes occur without the express knowledge of speakers and produce large-scale effects \cite{Keller1997}. Some of these effects have recently been the subject of quantitative studies \cite{Newberry2015,Shalom2019} using massive databases currently at our disposal \cite{Michel2010,davies2010corpus}. Of particular interest here is the analysis of the largest corpus of books available \cite{googlengrams}, which unveiled the dominance of 16-years oscillations in the frequency of occurrence of words, mounted over slowly varying trends \cite{Montemurro2016} (Figure \ref{fig2}a).

To explore the linguistic information carried by trends and oscillations, in a previous study we splitted both components from each word and grouped them by similarity \cite{PardoPintos2022}. The result of this operation is that words with similar trends form groups that characterize sociocultural periods, as shown in Figure \ref{fig2}b. In this example, the trends show a maximum in the early 20\ts{th} century, putting together keywords of the post-industrial society. Other groups gather keywords of the industrial revolution, the world wars and the digital era, among many others. 
Instead, when the trends are disregarded and only the oscillatory components of the words are considered, communities of semantically related words emerge, as the chemistry group in Figure \ref{fig2}c. Other groups with similar oscillations are related to economy, law, the army, medicine, etc..  
Here, we refer to communities of words with similar trends as {\em keywords}, and groups of words with similar oscillations as {\em semantic fields}.

The observation of regular oscillations led us to propose a basic mechanism for word usage \cite{PardoPintos2022} that is common to other cultural objects with life cycles, such as fashion \cite{Bikhchandani1992,Acerbi2012}. According to this interpretation, the words that belong to a field of interest increase their occurrences until sustained consumption produces a saturation that decreases usage, eventually leaving the topic ready to regain attention. We made this mechanism operative with the Volterra logistic model \cite{Li2004, Cheng2016}

\begin{equation} \label{eq:integro-diff}
\dot u=Ru\left[1-{\frac{1}{k}} \int_{-\infty}^t G(t-\tau)u(\tau)\,d\tau\right],
\end{equation}

\noindent where $R$ is the rate of growth of a word, and $G(t)$ a weighting factor that indicates how much emphasis should be given to the earlier times to determine its inhibitory effect in the present. We used the strong kernel $G(\tau)=4\tau/\bar\tau^2\ e^{-2\tau/\bar\tau}$, a distributed delay with a maximum influence $\bar\tau/2$ years in the past. With this kernel, equation \ref{eq:integro-diff}   can be rewritten as the 3-dimensional system \cite{fargue1973reducibilite}

\begin{equation} \label{eq:3dim}
    \left\{
        \begin{array}{ll}
          \dot u=Ru\,(1-w/k)\\
          \dot v=2/\bar\tau\,(u-v)\\
          \dot w=2/\bar\tau\,(v-w),
        \end{array}
    \right.
\end{equation}

\noindent which is indeed capable of producing cycles. In fact, a standard bifurcation analysis \cite{Lahcen2006} revealed that a stable periodic solution is created through a Hopf bifurcation as the delay is increased above the critical value $\bar\tau=4/R$ (Figure \ref{fig1}a).

The equilibrium $u^*=k$ represents the static usage of a word, a theoretical limit that is perturbed by the sociocultural context, identified with the experimental trends. Although simple, the  model of equations \ref{eq:3dim} driven by trends $k(t)$ fits the time series of word usage \cite{PardoPintos2022}, with parameters $R$ and $\bar\tau$ distributed along the Hopf bifurcation, as shown in Figure \ref{fig1}a. According to this view, word usage is represented by externally driven units poised near the birth of self-sustaining oscillations.

Although this model captures some central aspects of word usage, it treats each word in isolation, and therefore fails to explain the organization of words in communities \cite{Loreto2011}. The high degree of oscillatory coherence of the semantic communities, as displayed in the example of Figure \ref{fig2}c, suggests that words are phase-coupled. We tested this hypothesis by connecting Equations \ref{eq:3dim} with a Kuramoto model.

\section*{Results}

\begin{figure*}
\includegraphics[width=13cm]{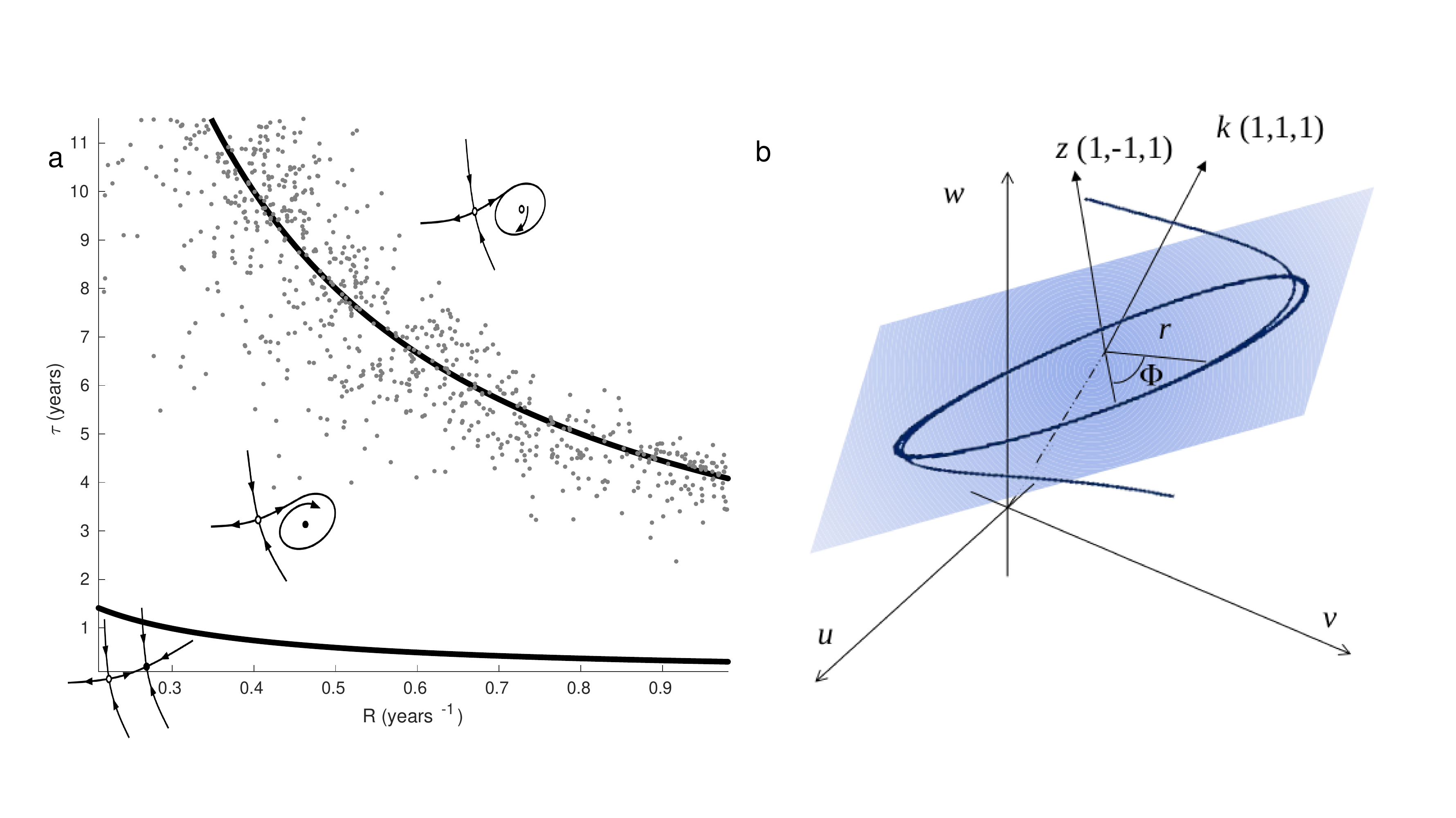}  
\caption{\textbf{Time series of word usage are fitted by a logistic model with strong kernel.} (a) Bifurcation diagram for Equation \ref{eq:3dim}. The origin is a saddle node. The other equilibrium, $(u,v,w)^*=k(1,1,1)$, undergoes a Hopf bifurcation at $\bar\tau=4/R$ (upper black line). The oscillations become increasingly damped until they disappear at $\bar\tau=8/27R$ (lower black line). The dimension not shown is attractive across the parameter space. English nouns fitted by the model are shown as grey points (mean growth rate $R=0.5\pm0.2$ years$^{-1}$ and mean delay $\bar\tau=8\pm3$ years). Simulations were performed by selecting paramters randomly from an area similar to the fitted series. This region was fitted to match the main coherence of the shuffled series (Figure \ref{fig2}) (b). In the Hopf bifurcation, a limit cycle is created in the plane of normal $(1,-1,1)$. The system can be rewritten in cylindrical coordinates $(r,\phi,z)$ that allows connecting the units of a word community through the  phase variable $\phi$.}
\label{fig1}
\end{figure*}

To map the system of equations \ref{eq:3dim} into a phase model \cite{Galan2005}, we translate the positive equilibrium $(u,v,w)^*=k(1,1,1)$ to the origin. In the region $\bar\tau>8/27R$ (above the lower curve in Figure \ref{fig1}), the linear part has two complex roots, $\Lambda$ and $\bar\Lambda$ and a real negative root, $\Lambda_{Re}$. Changing coordinates to the basis of eigenvectors and expressing the system in cylindrical coordinates yields

\begin{figure*}
\includegraphics[width=\textwidth]{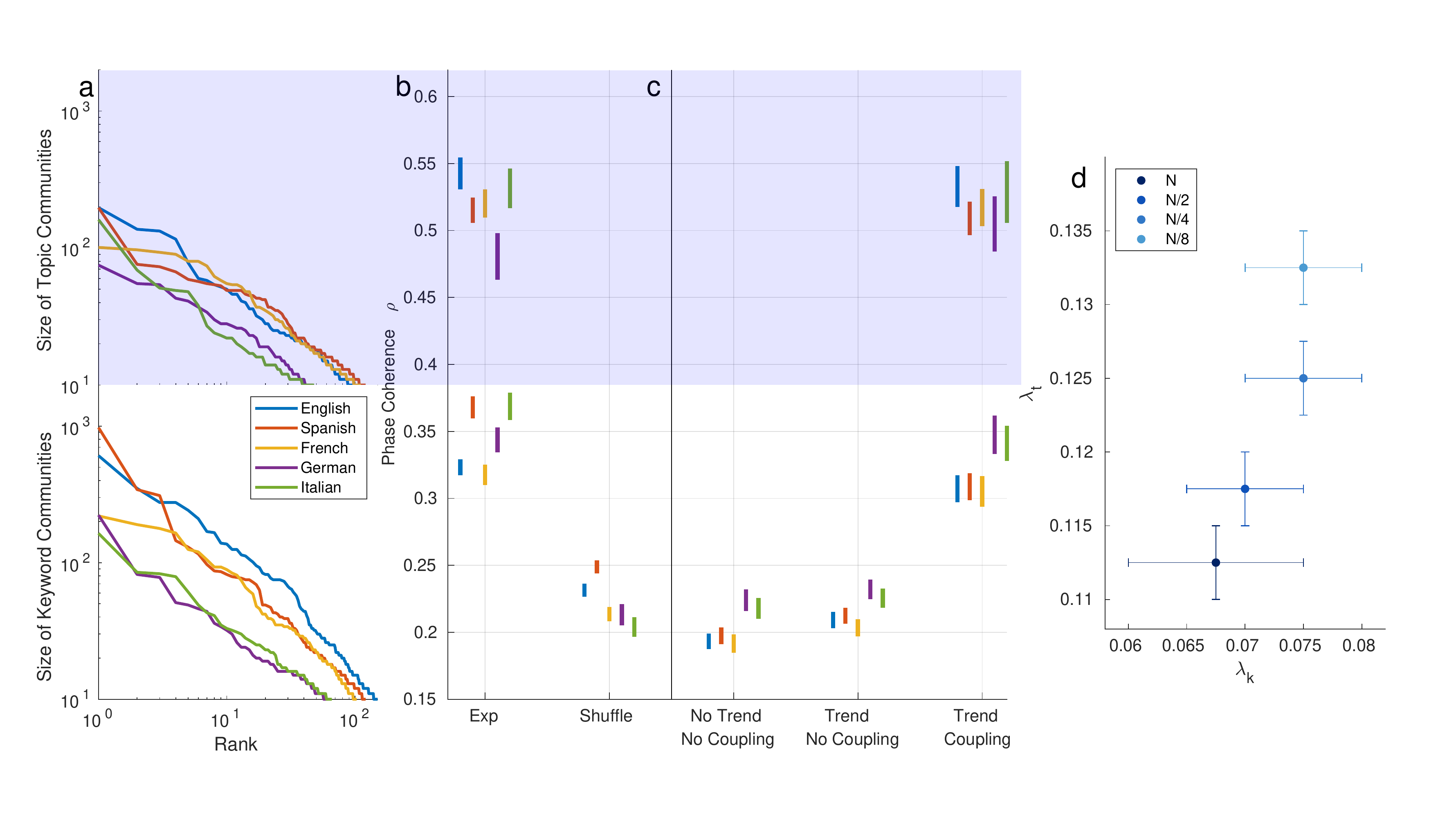}  
\caption{\textbf{Phase-coupled logistic equations account for the coherence observed in word communities.} (a) Topic and keyword communities ranked by size. (b) Mean coherence $\rho$ across communities for experimental data. (c) Mean coherence $\rho$ across communities for simulated data. Simulations of isolated units governed by Equation \ref{eq:3dim} with 
constant $k$ (no trend no coupling) are less coherent than the same units driven by the experimental trends $k=k(t)$ (trend no coupling). Weakly coupled driven units (trend coupling) exhibit coherence levels that are compatible with the experimental data. (d) Values of $\lambda_s$ and $\lambda_w$ for which simulations with coupling according to equation \ref{eq:kuramoto2} reach levels of coherence similar to the experimental ones when considering the network of all the words. The errorbars correspond to the step of the explored grid.}
\label{fig3}
\end{figure*}

\begin{equation}
\begin{split}
    \begin{pmatrix}
    \dot{r}\\ \dot{\phi}\\ \dot{z}
    \end{pmatrix}
    = &
R_z(\phi)
    \begin{pmatrix}
        \operatorname{Re}(\Lambda) & -\operatorname{Im}(\Lambda) & 0\\ \operatorname{Im}(\Lambda) & \operatorname{Re}(\Lambda) & 0 \\ 0 & 0 & \Lambda_{Re}
    \end{pmatrix}
    \begin{pmatrix}
        r\cos\phi \\ \sin\phi \\ z
    \end{pmatrix}+\\
    &+R_z(\phi) \begin{pmatrix}
    \operatorname{Re}(nl_1) \\ \operatorname{Im}(nl_1)/r \\ nl_3
    \end{pmatrix}
    \end{split}
    \label{eq:polar}
\end{equation}

\noindent where $R_z(\phi)$ is the rotation matriz along the direction $z=(1,-1,1)$, as sketched in Figure \ref{fig1}b; $T$ is the matrix of eigenvectors and $nl_{1,3}$ are the transformed nonlinear terms of equation \ref{eq:3dim} (see Methods). This allows us to represent a community $\mathpzc{N}$ of $N$ words by an equal number of systems of Equations \ref{eq:polar}, adding an all-to-all, purely sinusoidal Kuramoto coupling to the $\phi_i$ variable \cite{Kuramoto1984}

\begin{equation}
    \frac{\lambda}{N}\sum_{j\in \mathpzc{N}}\sin(\phi_j-\phi_i),
\label{eq:kuramoto}
\end{equation}

\noindent where $\lambda$ is a global coupling weight. 
Is this model capable of reproducing the coherence observed for semantic fields and keywords? To answer this question, we first begin by characterizing our experimental data. In Figure \ref{fig3}a, we show the distribution of the communities ranked by size, and in Figure \ref{fig3}b-Exp, we show the coherence $\bar\rho$ averaged across communities for semantic fields and keywords (see Methods).
By construction, semantic fields exhibit highly synchronized oscillations with a coherence of $\bar\rho\sim0.5$ across languages. As expected, the coherence of the keywords was lower, $\bar\rho\sim0.35$ (Figure \ref{fig3}b-Exp). Part of this coherence stems from finite size effects; to estimate this contribution, we shuffled the words between communities and recomputed the order parameter to obtain a baseline of $\bar\rho\sim0.2$ (Figure \ref{fig3}b-Shuffled).

To test the ability of our model to reproduce these properties, we simulated each community of $N$ words with an equal number of nodes controlled by Equations \ref{eq:3dim}. The initial conditions $(u_0,v_0,w_0)$ and parameters $(R,\bar\tau)$ were selected at random, the latter within the region with grey points around the Hopf bifurcation shown in Figure \ref{fig1}a. 
We then integrated the system expressed in cylindrical coordinates (Equations \ref{eq:polar}) driven by experimental trends and phase-coupled with a global weigth $\lambda$ (Equation \ref{eq:kuramoto}). We then invert the map to recover the usage $u(t)$ of all words in the community, over which we computed the order parameter $\rho$. Repeating this procedure across communities gives us the mean order parameter $\bar\rho$ (see Methods section).

Our simulations show that communities of nodes connected with a global coupling weight  $\lambda_t = 0.11$ years$^{-1}$ produce a mean coherence compatible with that observed for semantic fields across languages, $\bar\rho\sim0.5$ (Figure \ref{fig3}c-TrendCoupling). 
The keywords are perhaps more interesting because they allow the exploration of the relative roles of external driving and coupling. To see this, we begin with uncoupled non-driven nodes, for which we obtain coherence levels similar to those of shuffled words (Figure \ref{fig3}c-NoTrendNoCoupling). When the external drive is turned on, coherence increases (Figure \ref{fig3}c-TrendNoCoupling). This reflects the fact that collectively driven oscillators can be partially synchronized without being currently coupled \cite{Izhi, Levnajic2010}. Although the effect is slight, this shows that external driving indeed contributes to the coherence of the keyword communities  \footnote{In \cite{PardoPintos2022} we forced the units with a single representative trend of the community and found that the coherence increased to the levels observed experimentally.}. In fact, when the coupling is turned on, the coherence of keywords across languages ($\bar\rho\sim0.35$) is reached for $\lambda_k= 0.06$ 
 years$^{-1}$ (Figure \ref{fig3}c-TrendCoupling). 

Thus far, we have explored semantic fields and keywords separatedly. In each case, we showed that word communities can be described as nodes of  slowly driven  logistic equations linked by a Kuramoto coupling. Semantic fields and keywords differ only in their coupling weight, with $\lambda_t\sim2\lambda_k$.

We then address the problem of a complete network. By construction, the  communities of semantic fields  are mutually exclusive (a noun  does not belong to more than one topic), and the same applies to communities of keywords. However, semantic fields and keywords are not independent of each other, as illustrated by the word {\em copper} in Figure \ref{fig2}, which belongs to both the chemistry topic and the keywords of the early 20\ts{th} century. Hence, building a global picture of a complete network requires mixing both community types. This is performed using the following coupling  for the complete network:

\begin{equation}
    \frac{\lambda_k}{\kappa_{k_i}}\sum_{j\in \mathpzc{N}_{\,\,k}}\sin(\phi_j-\phi_i) + \frac{\lambda_k}{\kappa_{t_i}}\sum_{j\in \mathpzc{N}_{\,\,t}}\sin(\phi_j-\phi_i),
\label{eq:kuramoto2}
\end{equation}

\noindent where $\mathpzc{N}_{\,\,t}$ ($\mathpzc{N}_{\,\,k}$) is the topic (keyword) community to which the word $i$ belongs and $\kappa_i$ is the degree (number of links) of the node $i$ in each community.  
The dark blue dot in Figure \ref{fig3}d shows the coupling weights $\lambda_t$ and $\lambda_k$ that reproduce the experimental coherence. As expected, the mixture of communities required slightly higher values than those found when keywords and semantic fields were considered separatedly.

Finally, we explored the robustness of our results with respect to the network topology. For this sake, we relaxed the fully connected condition, simulating communities of $N$ words using $N/2$, $N/4$ and $N/8$ random connections while ensuring a finite distance between any two nodes. In Figure \ref{fig3}d  we see that when the number of connections within a community decreases, both coupling weigths increase to reach the experimental coherence level.

To summarize,  we show that the dynamics of word usage can be modelled as communities of phase-coupled logistic equations with distributed delays. 

\section*{Conclusions}

Logistic equations with distributed delays have a long-standing tradition in mathematical biology \cite{Lahcen2006}. Over the years, this bare-bones model has also proven useful for a variety of social phenomena, such as the upheavals of popular content in social media \cite{Lorenz-Spreen2019} and the cycles in word usage, interpreted as the interplay between interest and saturation \cite{PardoPintos2022}.  

Here we deal with the synchrony observed in different groups of words that tend to oscillate together in-phase. To account for this, we derived a phase map of the system, and linked the units with a Kuramoto coupling. Our main finding is that the oscillatory activity of the model reaches the coherence observed in the experimental communities, regardless of their topology and using a single global coupling for all the analyzed languages. 
   
The Kuramoto model relates in a very simple way to the original variables of the logistic equation. In fact, for a linear coupling of the state vectors, the corresponding phase function does not contain second or higher harmonics, as in the Kuramoto model \cite{Kori2014}.  According to this, the change in the occurrence of a word depends linearly on the occurrence of its semantic neighbours. This simple relationship between the words of a given semantic field allows us to increase confidence in this model for word usage dynamics.

A final remark regarding the scale of description of the problem. In this study we focus on the nouns present throughout the last three centuries, excluding from the analysis the words that enter or leave the corpus in that period and the changes in the word clusters that may occur when the network is analyzed at a higher temporal resolution. Here we describe the network associated with the long-term dynamics of word usage, a necessary first step before describing the dynamics of clusters at a higher resolution, for which the present description in the phase domain offers a vast battery of analysis tools \cite{Ashwin2016,gutierrez2011}.

\section*{Materials and Methods}
\label{sec:Methods}


\noindent{\bf Data processing.} Google Books is a massive corpus of lexical data extracted from approximately eight million books (6\% of all books ever published) that has been widely used for research. Despite its size, the database is not free from biases \cite{Pechenick2015}, which we addressed in \cite{PardoPintos2022}.

Briefly, we collected tokens  of the most common nouns converted to singular forms in English (10,403), Spanish (8,064), French (6,291), German (3,341), and Italian (2,995) from Google Books 2019 \cite{Michel2011}, retaining only nouns with at least $10^6$ appearances per year over the last 300 years. 
We then computed the word frequency $x(t)=n(t)/N(t)$, where $n(t)$ is the number of appearances of noun $n$, and $N(t)$ is the size of the corpus in year $t$. Singular spectral analysis (SSA) \cite{Zhigljavsky2011} was used to extract the trends $k(t)$, computed as the non-cyclic components of the time series $x(t)$ \cite{alexandrov2005automatic}. The oscillatory components $x(t)-k(t)$ were low-pass filtered ($f<1/6$ years$^{-1}$) to avoid possible random sampling effects in database loading \cite{Karjus2020}.

\vspace{.5cm}
\noindent{\bf Clustering.} The semantic fields and keywords were computed from the correlation matrices of trends $k(t)$ and oscillations $x(t)-k(t)$ for the extracted nouns in English, Spanish, German, French, and Italian. The  communities were determined using a cutoff for the correlations of 0.04 for trends and 0.5 for oscillations; these values ensure maximum correlation between series compatible with a cluster size distribution that follows Zipf's law. Communities with fewer than 10 words were discarded.


\vspace{.5cm}
\noindent{\bf Phase coherence.} We transformed the oscillations $x(t)-k(t)$ into phase variables $\theta(t)$ using the Hilbert transform \cite{Matsuda2017}. The collective rhythm of a community of $N$ words was then computed using the order parameter averaged over the last three centuries, $\rho=\left\langle|\sum_{j=1}^N e^{i\theta_j(t)}/N |\right\rangle$. The mean coherence across communities, $\bar\rho$ was then computed.
Figure \ref{fig3} shows the distribution of mean coherence values across communities for different languages. Coherence values were normally distributed (Kolmogorov-Smirnov test: $p>0.05$ for all languages and conditions). A two-sample t-test showed that only the distribution of experimental data (Fig. \ref{fig3}a keywords) and the simulations of coupled externally driven units (Figure \ref{fig3}b trend coupling) are equivalent across languages.

\vspace{.5cm}
\noindent{\bf Derivation of the phase model.} Translating the equations \ref{eq:3dim} to the slowly evolving point $(k,k,k)$ we have



\begin{equation*}
    \begin{pmatrix}
    \dot{u}\\ \dot{v}\\ \dot{w}
    \end{pmatrix}
    = 
    \begin{pmatrix}
        0 & 0 & -R\\2/\bar\tau & -2/\bar\tau & 0 \\ 0 & 2/\bar\tau & -2/\bar\tau
    \end{pmatrix}
    \begin{pmatrix}
        u \\ v \\ w
    \end{pmatrix}
    -
    \begin{pmatrix}
        \frac{R}{k}uw+\dot{k} \\ \dot{k} \\ \dot{k}
    \end{pmatrix}.
\end{equation*}

Considering the linear part, we obtain the characteristic equation
$\Lambda^3+4/\bar\tau\,\Lambda^2+4/\bar\tau^2\,\Lambda+4R/\bar\tau^2=0$. For $\bar\tau>8/27R$ (above the lower curve in Figure \ref{fig1}), we have two complex conjugate roots $\Lambda_1$ and $\Lambda_2$ and a real negative root $ \Lambda_3$:

\begin{equation*}
\begin{split}
&\Lambda_1 = \frac{1}{3\bar\tau} \left( L_1 + L_2 - 4 \right) \\
&\Lambda_{2,3} = \frac{1}{3\bar\tau} \left[ -\frac{1}{4} \left( L_1 + L_2 \right) - 4 \pm j \frac{\sqrt{3}}{2}  \left( L_2 - L_1\right) \right],\\
\end{split}
\end{equation*}

\noindent where $L_1 = 2^{1/3}C^{1/3}$,  $L_2 =2^{5/3} C^{-1/3}$, and $C = 3 \sqrt{3R\bar\tau}\sqrt{27R\bar\tau-8}-27R\bar\tau +4$  is complex. Changing coordinates to the basis of eigenvectors, the  system reads

\begin{equation}
 \begin{split}
    &\begin{pmatrix}
    \dot{x_1}\\ \dot{x_2}\\ \dot{x_3}
    \end{pmatrix}
     = 
    \begin{pmatrix}
        \Lambda_1 & 0 & 0\\ 0 & \Lambda_2 & 0 \\ 0 & 0 & \Lambda_3
    \end{pmatrix}
    \begin{pmatrix}
        x_1 \\ x_2 \\ x_3
    \end{pmatrix}-
    T^{-1} \begin{pmatrix}
        \frac{R}{k} \sum\limits_{i,j} x_ix_jA_i^2 +\dot{k}  \\ \dot{k}\\  \dot{k}
    \end{pmatrix}
    \end{split}
    \label{eq:complex}
\end{equation}

\noindent with $A_i = 1+\Lambda_i\bar\tau/2$, $x_1$ and $x_2$  complex conjugate variables and $T$ the matrix of eigenvectors
\begin{equation*}
\begin{split}
    T &= \begin{pmatrix}
        A_1^2 & A_2^2 & A_3^2 \\
        A_1 & A_2 & A_3 \\
        1 & 1 & 1
        \end{pmatrix}. \\
\end{split}
\end{equation*}

This system can be expressed as:

\begin{equation}
\begin{split}
    \begin{pmatrix}
    \dot{r}\\ \dot{\phi}\\ \dot{z}
    \end{pmatrix}
    = &
R_z(\phi)
    \begin{pmatrix}
        \operatorname{Re}(\Lambda_1) & -\operatorname{Im}(\Lambda_1) & 0\\ \operatorname{Im}(\Lambda_1) & \operatorname{Re}(\Lambda_1) & 0 \\ 0 & 0 & \Lambda_3
    \end{pmatrix}
    \begin{pmatrix}
        r\cos\phi \\ \sin\phi \\ z
    \end{pmatrix}+\\
    &+R_z(\phi) \begin{pmatrix}
    \operatorname{Re}(nl_1) \\ \operatorname{Im}(nl_1/r) \\ 
    nl_3
    \end{pmatrix}
    \end{split}
    \label{eq:cylindrical}
  \end{equation}

\noindent where $nl_1$ and $nl_3$ stand for the nonlinear terms of the $\dot{x}_1$ and $\dot{x}_3$ in equation \ref{eq:complex},

\begin{equation*}
\begin{split}
&nl_1 = \frac{L_1^2(L_1^2-4)}{18R\bar\tau D} \cdot \left( \frac{R}{k}\sum\limits_{i,j} x_ix_jA_i^2 +\dot{k} \right) - \\
&-\frac{\dot k}{3L_1^2} \left[2(C+2+B-L_1^2)+\sqrt{3}j(C+2-B)\sum\limits_{i=1}^{2}(-1)^{i+1}\right],\\
    \end{split}
\end{equation*}


\begin{equation*}
\begin{split}
&nl_3 = \frac{L_1^2 \left[-(L_1^2-4)+\sqrt{3}j(L_1^2+4)\right]}{36R\bar\tau D} \left( \frac{R}{k}\sum\limits_{i,j} x_ix_jA_i^2 +\dot{k} \right)\\
&+ \frac{L_1 \left[ (C+2-B)-\sqrt{3}j (C+2+B)\right]}{27R\bar\tau D} \dot{k}+ \\
&+ \left[1+\frac{(L_1^5/2-4L_1^2+E)+\sqrt{3}j(L_1^5/2-4L_1^2-E)}{108R\bar\tau D} \right] \dot{k}
    \end{split}
\end{equation*}

\noindent where $B = L_1^4/8+2L_1$, $D = 4 \sqrt{27R\bar\tau -8}/(3 \sqrt{3R\bar\tau})-4-C$ and $E =L_1^4-8L_1$. The matrix $R_z(\phi)$ is the usual rotation matrix along the $z$ axis

\begin{equation*}
R_z(\phi) = \begin{pmatrix}
        \cos\phi & -\sin\phi & 0\\ \sin\phi & \cos\phi & 0 \\ 0 & 0 & 1
    \end{pmatrix}.
\end{equation*}

Equation \ref{eq:cylindrical} represents a phase map of the slowly driven logistic equation with a strong kernel across the entire oscillatory region above the lower curve $\bar\tau=8/(27R)$ in Figure \ref{fig1}a. This expression was combined with the Kuramoto phase-coupling equation \ref{eq:kuramoto} to model the communities of words.


\vspace{.5cm}
\noindent{\bf Data and code availability.} 
All the datasets are publicly available at \footnote{  \href{http://storage.googleapis.com/books/ngrams/books/datasetsv2.html}{http://storage.googleapis.com/books/ngrams/books/} datasetsv2.html}. The matlab codes and processed data used to generate the Figures of this work and the word communities are available at \footnote{ \href{https://github.com/AlePardoPintos/Words-Phase-Coupled}{https://github.com/AlePardoPintos/Words-Phase-Coupled}}.

\vspace{.5cm}

\begin{acknowledgments}
This study was partially funded by the University of Buenos Aires (UBA) and the Consejo Nacional de Investigaciones Científicas y Técnicas (CONICET) through the grant PIP-11220200102083CO.
\end{acknowledgments}

\vspace{6cm}

\bibliography{MyCollection}

\begin{thebibliography}{32}%
\makeatletter
\providecommand \@ifxundefined [1]{%
 \@ifx{#1\undefined}
}%
\providecommand \@ifnum [1]{%
 \ifnum #1\expandafter \@firstoftwo
 \else \expandafter \@secondoftwo
 \fi
}%
\providecommand \@ifx [1]{%
 \ifx #1\expandafter \@firstoftwo
 \else \expandafter \@secondoftwo
 \fi
}%
\providecommand \natexlab [1]{#1}%
\providecommand \enquote  [1]{``#1''}%
\providecommand \bibnamefont  [1]{#1}%
\providecommand \bibfnamefont [1]{#1}%
\providecommand \citenamefont [1]{#1}%
\providecommand \href@noop [0]{\@secondoftwo}%
\providecommand \href [0]{\begingroup \@sanitize@url \@href}%
\providecommand \@href[1]{\@@startlink{#1}\@@href}%
\providecommand \@@href[1]{\endgroup#1\@@endlink}%
\providecommand \@sanitize@url [0]{\catcode `\\12\catcode `\$12\catcode
  `\&12\catcode `\#12\catcode `\^12\catcode `\_12\catcode `\%12\relax}%
\providecommand \@@startlink[1]{}%
\providecommand \@@endlink[0]{}%
\providecommand \url  [0]{\begingroup\@sanitize@url \@url }%
\providecommand \@url [1]{\endgroup\@href {#1}{\urlprefix }}%
\providecommand \urlprefix  [0]{URL }%
\providecommand \Eprint [0]{\href }%
\providecommand \doibase [0]{https://doi.org/}%
\providecommand \selectlanguage [0]{\@gobble}%
\providecommand \bibinfo  [0]{\@secondoftwo}%
\providecommand \bibfield  [0]{\@secondoftwo}%
\providecommand \translation [1]{[#1]}%
\providecommand \BibitemOpen [0]{}%
\providecommand \bibitemStop [0]{}%
\providecommand \bibitemNoStop [0]{.\EOS\space}%
\providecommand \EOS [0]{\spacefactor3000\relax}%
\providecommand \BibitemShut  [1]{\csname bibitem#1\endcsname}%
\let\auto@bib@innerbib\@empty
\bibitem [{\citenamefont {Keller}(1997)}]{Keller1997}%
  \BibitemOpen
  \bibfield  {author} {\bibinfo {author} {\bibfnamefont {R.}~\bibnamefont
  {Keller}},\ }\href {https://doi.org/10.2307/416617} {\emph {\bibinfo {title}
  {Language}}},\ Vol.~\bibinfo {volume} {73}\ (\bibinfo  {publisher}
  {Routledge},\ \bibinfo {address} {London and New York},\ \bibinfo {year}
  {1997})\ p.\ \bibinfo {pages} {193}\BibitemShut {NoStop}%
\bibitem [{\citenamefont {Newberry}\ \emph {et~al.}(2015)\citenamefont
  {Newberry}, \citenamefont {Ennis},\ and\ \citenamefont
  {Savage}}]{Newberry2015}%
  \BibitemOpen
  \bibfield  {author} {\bibinfo {author} {\bibfnamefont {M.~G.}\ \bibnamefont
  {Newberry}}, \bibinfo {author} {\bibfnamefont {D.~B.}\ \bibnamefont
  {Ennis}},\ and\ \bibinfo {author} {\bibfnamefont {V.~M.}\ \bibnamefont
  {Savage}},\ }\href {https://doi.org/10.1371/journal.pcbi.1004455} {\bibfield
  {journal} {\bibinfo  {journal} {PLoS Computational Biology}\ }\textbf
  {\bibinfo {volume} {11}},\ \bibinfo {pages} {1} (\bibinfo {year}
  {2015})}\BibitemShut {NoStop}%
\bibitem [{\citenamefont {Shalom}\ \emph {et~al.}(2019)\citenamefont {Shalom},
  \citenamefont {Sigman}, \citenamefont {Mindlin},\ and\ \citenamefont
  {Trevisan}}]{Shalom2019}%
  \BibitemOpen
  \bibfield  {author} {\bibinfo {author} {\bibfnamefont {D.~E.}\ \bibnamefont
  {Shalom}}, \bibinfo {author} {\bibfnamefont {M.}~\bibnamefont {Sigman}},
  \bibinfo {author} {\bibfnamefont {G.}~\bibnamefont {Mindlin}},\ and\ \bibinfo
  {author} {\bibfnamefont {M.~A.}\ \bibnamefont {Trevisan}},\ }\href
  {https://doi.org/10.1103/PhysRevE.100.020102} {\bibfield  {journal} {\bibinfo
   {journal} {Physical Review E}\ }\textbf {\bibinfo {volume} {100}},\ \bibinfo
  {pages} {20102} (\bibinfo {year} {2019})},\ \Eprint
  {https://arxiv.org/abs/1811.08465} {arXiv:1811.08465} \BibitemShut {NoStop}%
\bibitem [{\citenamefont {Michel}\ \emph {et~al.}(2010)\citenamefont {Michel},
  \citenamefont {Shen}, \citenamefont {Aiden}, \citenamefont {Veres},
  \citenamefont {Gray}, \citenamefont {Pickett}, \citenamefont {Hoiberg},
  \citenamefont {Clancy}, \citenamefont {Norvig}, \citenamefont {Orwant},
  \citenamefont {Pinker}, \citenamefont {Nowak},\ and\ \citenamefont
  {Aiden}}]{Michel2010}%
  \BibitemOpen
  \bibfield  {author} {\bibinfo {author} {\bibfnamefont {J.-B.}\ \bibnamefont
  {Michel}}, \bibinfo {author} {\bibfnamefont {Y.~K.}\ \bibnamefont {Shen}},
  \bibinfo {author} {\bibfnamefont {a.~P.}\ \bibnamefont {Aiden}}, \bibinfo
  {author} {\bibfnamefont {A.}~\bibnamefont {Veres}}, \bibinfo {author}
  {\bibfnamefont {M.~K.}\ \bibnamefont {Gray}}, \bibinfo {author}
  {\bibfnamefont {J.~P.}\ \bibnamefont {Pickett}}, \bibinfo {author}
  {\bibfnamefont {D.}~\bibnamefont {Hoiberg}}, \bibinfo {author} {\bibfnamefont
  {D.}~\bibnamefont {Clancy}}, \bibinfo {author} {\bibfnamefont
  {P.}~\bibnamefont {Norvig}}, \bibinfo {author} {\bibfnamefont
  {J.}~\bibnamefont {Orwant}}, \bibinfo {author} {\bibfnamefont
  {S.}~\bibnamefont {Pinker}}, \bibinfo {author} {\bibfnamefont {M.~a.}\
  \bibnamefont {Nowak}},\ and\ \bibinfo {author} {\bibfnamefont {E.~L.}\
  \bibnamefont {Aiden}},\ }\href {https://doi.org/10.1126/science.1199644}
  {\bibfield  {journal} {\bibinfo  {journal} {Science}\ }\textbf {\bibinfo
  {volume} {331}},\ \bibinfo {pages} {176} (\bibinfo {year}
  {2010})}\BibitemShut {NoStop}%
\bibitem [{\citenamefont {Davies}(2010)}]{davies2010corpus}%
  \BibitemOpen
  \bibfield  {author} {\bibinfo {author} {\bibfnamefont {M.}~\bibnamefont
  {Davies}},\ }\href@noop {} {\emph {\bibinfo {title} {The corpus of historical
  American English: COHA}}}\ (\bibinfo  {publisher} {BYE, Brigham Young
  University},\ \bibinfo {year} {2010})\BibitemShut {NoStop}%
\bibitem [{goo(2019)}]{googlengrams}%
  \BibitemOpen
  \href {https://books.google.com/ngrams/} {\bibinfo {title} {{Google Books
  Ngram viewer}}} (\bibinfo {year} {2019})\BibitemShut {NoStop}%
\bibitem [{\citenamefont {Montemurro}\ and\ \citenamefont
  {Zanette}(2016)}]{Montemurro2016}%
  \BibitemOpen
  \bibfield  {author} {\bibinfo {author} {\bibfnamefont {M.~A.}\ \bibnamefont
  {Montemurro}}\ and\ \bibinfo {author} {\bibfnamefont {D.~H.}\ \bibnamefont
  {Zanette}},\ }\href {https://doi.org/10.1057/palcomms.2016.84} {\bibfield
  {journal} {\bibinfo  {journal} {Palgrave Communications}\ }\textbf {\bibinfo
  {volume} {2}},\ \bibinfo {pages} {1} (\bibinfo {year} {2016})}\BibitemShut
  {NoStop}%
\bibitem [{\citenamefont {{Pardo Pintos}}\ \emph {et~al.}(2022)\citenamefont
  {{Pardo Pintos}}, \citenamefont {Shalom}, \citenamefont {Tagliazucchi},
  \citenamefont {Mindlin},\ and\ \citenamefont {Trevisan}}]{PardoPintos2022}%
  \BibitemOpen
  \bibfield  {author} {\bibinfo {author} {\bibfnamefont {A.}~\bibnamefont
  {{Pardo Pintos}}}, \bibinfo {author} {\bibfnamefont {D.~E.}\ \bibnamefont
  {Shalom}}, \bibinfo {author} {\bibfnamefont {E.}~\bibnamefont
  {Tagliazucchi}}, \bibinfo {author} {\bibfnamefont {G.}~\bibnamefont
  {Mindlin}},\ and\ \bibinfo {author} {\bibfnamefont {M.}~\bibnamefont
  {Trevisan}},\ }\href {https://doi.org/10.1016/j.chaos.2022.112327} {\bibfield
   {journal} {\bibinfo  {journal} {Chaos, Solitons and Fractals}\ }\textbf
  {\bibinfo {volume} {161}},\ \bibinfo {pages} {1} (\bibinfo {year} {2022})},\
  \Eprint {https://arxiv.org/abs/2201.04739} {arXiv:2201.04739} \BibitemShut
  {NoStop}%
\bibitem [{\citenamefont {Bikhchandani}\ \emph {et~al.}(1992)\citenamefont
  {Bikhchandani}, \citenamefont {Hirshleifer},\ and\ \citenamefont
  {Welch}}]{Bikhchandani1992}%
  \BibitemOpen
  \bibfield  {author} {\bibinfo {author} {\bibfnamefont {S.}~\bibnamefont
  {Bikhchandani}}, \bibinfo {author} {\bibfnamefont {D.}~\bibnamefont
  {Hirshleifer}},\ and\ \bibinfo {author} {\bibfnamefont {I.}~\bibnamefont
  {Welch}},\ }\href {https://doi.org/10.1086/261849} {\bibfield  {journal}
  {\bibinfo  {journal} {Journal of Political Economy}\ }\textbf {\bibinfo
  {volume} {100}},\ \bibinfo {pages} {992} (\bibinfo {year} {1992})},\ \Eprint
  {https://arxiv.org/abs/arXiv:1011.1669v3} {arXiv:arXiv:1011.1669v3}
  \BibitemShut {NoStop}%
\bibitem [{\citenamefont {Acerbi}\ \emph {et~al.}(2012)\citenamefont {Acerbi},
  \citenamefont {Ghirlanda},\ and\ \citenamefont {Enquist}}]{Acerbi2012}%
  \BibitemOpen
  \bibfield  {author} {\bibinfo {author} {\bibfnamefont {A.}~\bibnamefont
  {Acerbi}}, \bibinfo {author} {\bibfnamefont {S.}~\bibnamefont {Ghirlanda}},\
  and\ \bibinfo {author} {\bibfnamefont {M.}~\bibnamefont {Enquist}},\
  }\bibfield  {journal} {\bibinfo  {journal} {PLoS ONE}\ }\textbf {\bibinfo
  {volume} {7}},\ \href {https://doi.org/10.1371/journal.pone.0032541}
  {10.1371/journal.pone.0032541} (\bibinfo {year} {2012})\BibitemShut {NoStop}%
\bibitem [{\citenamefont {Li}\ \emph {et~al.}(2004)\citenamefont {Li},
  \citenamefont {Liao},\ and\ \citenamefont {Li}}]{Li2004}%
  \BibitemOpen
  \bibfield  {author} {\bibinfo {author} {\bibfnamefont {S.}~\bibnamefont
  {Li}}, \bibinfo {author} {\bibfnamefont {X.}~\bibnamefont {Liao}},\ and\
  \bibinfo {author} {\bibfnamefont {C.}~\bibnamefont {Li}},\ }\href
  {https://doi.org/10.1016/j.chaos.2004.02.048} {\bibfield  {journal} {\bibinfo
   {journal} {Chaos, Solitons and Fractals}\ }\textbf {\bibinfo {volume}
  {22}},\ \bibinfo {pages} {713} (\bibinfo {year} {2004})}\BibitemShut
  {NoStop}%
\bibitem [{\citenamefont {Cheng}\ \emph {et~al.}(2016)\citenamefont {Cheng},
  \citenamefont {Wang},\ and\ \citenamefont {Cao}}]{Cheng2016}%
  \BibitemOpen
  \bibfield  {author} {\bibinfo {author} {\bibfnamefont {Z.}~\bibnamefont
  {Cheng}}, \bibinfo {author} {\bibfnamefont {Y.}~\bibnamefont {Wang}},\ and\
  \bibinfo {author} {\bibfnamefont {J.}~\bibnamefont {Cao}},\ }\href
  {https://doi.org/10.1007/s11071-016-2891-1} {\bibfield  {journal} {\bibinfo
  {journal} {Nonlinear Dynamics}\ }\textbf {\bibinfo {volume} {86}},\ \bibinfo
  {pages} {323} (\bibinfo {year} {2016})}\BibitemShut {NoStop}%
\bibitem [{\citenamefont {Fargue}(1973)}]{fargue1973reducibilite}%
  \BibitemOpen
  \bibfield  {author} {\bibinfo {author} {\bibfnamefont {D.}~\bibnamefont
  {Fargue}},\ }\href@noop {} {\bibfield  {journal} {\bibinfo  {journal} {CR
  Acad. Sci. Paris B}\ }\textbf {\bibinfo {volume} {277}},\ \bibinfo {pages}
  {471} (\bibinfo {year} {1973})}\BibitemShut {NoStop}%
\bibitem [{\citenamefont {Lahcen}(2006)}]{Lahcen2006}%
  \BibitemOpen
  \bibfield  {author} {\bibinfo {author} {\bibfnamefont {M.}~\bibnamefont
  {Lahcen}},\ }\href {https://doi.org/10.1007/1-4020-3647-7} {\emph {\bibinfo
  {title} {Delay Differential Equations and Applications}}},\ \bibinfo {number}
  {March}\ (\bibinfo  {publisher} {Springer},\ \bibinfo {year}
  {2006})\BibitemShut {NoStop}%
\bibitem [{\citenamefont {Loreto}\ \emph {et~al.}(2011)\citenamefont {Loreto},
  \citenamefont {Baronchelli}, \citenamefont {Mukherjee}, \citenamefont
  {Puglisi},\ and\ \citenamefont {Tria}}]{Loreto2011}%
  \BibitemOpen
  \bibfield  {author} {\bibinfo {author} {\bibfnamefont {V.}~\bibnamefont
  {Loreto}}, \bibinfo {author} {\bibfnamefont {A.}~\bibnamefont {Baronchelli}},
  \bibinfo {author} {\bibfnamefont {A.}~\bibnamefont {Mukherjee}}, \bibinfo
  {author} {\bibfnamefont {A.}~\bibnamefont {Puglisi}},\ and\ \bibinfo {author}
  {\bibfnamefont {F.}~\bibnamefont {Tria}},\ }\bibfield  {journal} {\bibinfo
  {journal} {Journal of Statistical Mechanics: Theory and Experiment}\ }\textbf
  {\bibinfo {volume} {2011}},\ \href
  {https://doi.org/10.1088/1742-5468/2011/04/P04006}
  {10.1088/1742-5468/2011/04/P04006} (\bibinfo {year} {2011})\BibitemShut
  {NoStop}%
\bibitem [{\citenamefont {Gal{\'{a}}n}\ \emph {et~al.}(2005)\citenamefont
  {Gal{\'{a}}n}, \citenamefont {Ermentrout},\ and\ \citenamefont
  {Urban}}]{Galan2005}%
  \BibitemOpen
  \bibfield  {author} {\bibinfo {author} {\bibfnamefont {R.~F.}\ \bibnamefont
  {Gal{\'{a}}n}}, \bibinfo {author} {\bibfnamefont {G.~B.}\ \bibnamefont
  {Ermentrout}},\ and\ \bibinfo {author} {\bibfnamefont {N.~N.}\ \bibnamefont
  {Urban}},\ }\href {https://doi.org/10.1103/PhysRevLett.94.158101} {\bibfield
  {journal} {\bibinfo  {journal} {Physical Review Letters}\ }\textbf {\bibinfo
  {volume} {94}},\ \bibinfo {pages} {1} (\bibinfo {year} {2005})}\BibitemShut
  {NoStop}%
\bibitem [{\citenamefont {Kuramoto}(1984)}]{Kuramoto1984}%
  \BibitemOpen
  \bibfield  {author} {\bibinfo {author} {\bibfnamefont {Y.}~\bibnamefont
  {Kuramoto}},\ }\href {https://doi.org/10.1007/978-3-642-12601-7} {\emph
  {\bibinfo {title} {Springer Series in Synergetics}}}\ (\bibinfo {year}
  {1984})\ p.\ \bibinfo {pages} {416},\ \Eprint
  {https://arxiv.org/abs/arXiv:1011.1669v3} {arXiv:arXiv:1011.1669v3}
  \BibitemShut {NoStop}%
\bibitem [{\citenamefont {Hoppensteadt}\ and\ \citenamefont
  {Izhikevich}(1997)}]{Izhi}%
  \BibitemOpen
  \bibfield  {author} {\bibinfo {author} {\bibfnamefont {F.~C.}\ \bibnamefont
  {Hoppensteadt}}\ and\ \bibinfo {author} {\bibfnamefont {E.~M.}\ \bibnamefont
  {Izhikevich}},\ }\href@noop {} {\emph {\bibinfo {title} {{Weakly connected
  neural networks}}}}\ (\bibinfo  {publisher} {Springer-Verlag},\ \bibinfo
  {address} {New York},\ \bibinfo {year} {1997})\BibitemShut {NoStop}%
\bibitem [{\citenamefont {Levnaji{\'{c}}}\ and\ \citenamefont
  {Pikovsky}(2010)}]{Levnajic2010}%
  \BibitemOpen
  \bibfield  {author} {\bibinfo {author} {\bibfnamefont {Z.}~\bibnamefont
  {Levnaji{\'{c}}}}\ and\ \bibinfo {author} {\bibfnamefont {A.}~\bibnamefont
  {Pikovsky}},\ }\href {https://doi.org/10.1103/PhysRevE.82.056202} {\bibfield
  {journal} {\bibinfo  {journal} {Physical Review E - Statistical, Nonlinear,
  and Soft Matter Physics}\ }\textbf {\bibinfo {volume} {82}},\ \bibinfo
  {pages} {1} (\bibinfo {year} {2010})},\ \Eprint
  {https://arxiv.org/abs/1007.4097} {arXiv:1007.4097} \BibitemShut {NoStop}%
\bibitem [{Note1()}]{Note1}%
  \BibitemOpen
  \bibinfo {note} {In \cite {PardoPintos2022} we forced the units with a single
  representative trend of the community and found that the coherence increased
  to the levels observed experimentally.}\BibitemShut {Stop}%
\bibitem [{\citenamefont {Lorenz-Spreen}\ \emph {et~al.}(2019)\citenamefont
  {Lorenz-Spreen}, \citenamefont {M{\o}nsted}, \citenamefont {H{\"{o}}vel},\
  and\ \citenamefont {Lehmann}}]{Lorenz-Spreen2019}%
  \BibitemOpen
  \bibfield  {author} {\bibinfo {author} {\bibfnamefont {P.}~\bibnamefont
  {Lorenz-Spreen}}, \bibinfo {author} {\bibfnamefont {B.~M.}\ \bibnamefont
  {M{\o}nsted}}, \bibinfo {author} {\bibfnamefont {P.}~\bibnamefont
  {H{\"{o}}vel}},\ and\ \bibinfo {author} {\bibfnamefont {S.}~\bibnamefont
  {Lehmann}},\ }\href {https://doi.org/10.1038/s41467-019-09311-w} {\bibfield
  {journal} {\bibinfo  {journal} {Nature Communications}\ }\textbf {\bibinfo
  {volume} {10}},\ \bibinfo {pages} {1} (\bibinfo {year} {2019})}\BibitemShut
  {NoStop}%
\bibitem [{\citenamefont {Kori}\ \emph {et~al.}(2014)\citenamefont {Kori},
  \citenamefont {Kuramoto}, \citenamefont {Jain}, \citenamefont {Kiss},\ and\
  \citenamefont {Hudson}}]{Kori2014}%
  \BibitemOpen
  \bibfield  {author} {\bibinfo {author} {\bibfnamefont {H.}~\bibnamefont
  {Kori}}, \bibinfo {author} {\bibfnamefont {Y.}~\bibnamefont {Kuramoto}},
  \bibinfo {author} {\bibfnamefont {S.}~\bibnamefont {Jain}}, \bibinfo {author}
  {\bibfnamefont {I.~Z.}\ \bibnamefont {Kiss}},\ and\ \bibinfo {author}
  {\bibfnamefont {J.~L.}\ \bibnamefont {Hudson}},\ }\bibfield  {journal}
  {\bibinfo  {journal} {Physical Review E - Statistical, Nonlinear, and Soft
  Matter Physics}\ }\textbf {\bibinfo {volume} {89}},\ \href
  {https://doi.org/10.1103/PhysRevE.89.062906} {10.1103/PhysRevE.89.062906}
  (\bibinfo {year} {2014})\BibitemShut {NoStop}%
\bibitem [{\citenamefont {Ashwin}\ and\ \citenamefont
  {Rodrigues}(2016)}]{Ashwin2016}%
  \BibitemOpen
  \bibfield  {author} {\bibinfo {author} {\bibfnamefont {P.}~\bibnamefont
  {Ashwin}}\ and\ \bibinfo {author} {\bibfnamefont {A.}~\bibnamefont
  {Rodrigues}},\ }\href {https://doi.org/10.1016/j.physd.2016.02.009}
  {\bibfield  {journal} {\bibinfo  {journal} {Physica D: Nonlinear Phenomena}\
  }\textbf {\bibinfo {volume} {325}},\ \bibinfo {pages} {14} (\bibinfo {year}
  {2016})}\BibitemShut {NoStop}%
\bibitem [{\citenamefont {Gutiérrez}\ \emph {et~al.}(2011)\citenamefont
  {Gutiérrez}, \citenamefont {Amann}, \citenamefont {Assenza}, \citenamefont
  {Gómez-Gardeñes}, \citenamefont {Latora},\ and\ \citenamefont
  {Boccaletti}}]{gutierrez2011}%
  \BibitemOpen
  \bibfield  {author} {\bibinfo {author} {\bibfnamefont {R.}~\bibnamefont
  {Gutiérrez}}, \bibinfo {author} {\bibfnamefont {A.}~\bibnamefont {Amann}},
  \bibinfo {author} {\bibfnamefont {S.}~\bibnamefont {Assenza}}, \bibinfo
  {author} {\bibfnamefont {J.}~\bibnamefont {Gómez-Gardeñes}}, \bibinfo
  {author} {\bibfnamefont {V.}~\bibnamefont {Latora}},\ and\ \bibinfo {author}
  {\bibfnamefont {S.}~\bibnamefont {Boccaletti}},\ }\href
  {https://doi.org/10.1103/PhysRevLett.107.234103} {\bibfield  {journal}
  {\bibinfo  {journal} {Physical Review Letters}\ }\textbf {\bibinfo {volume}
  {107}},\ \bibinfo {pages} {1} (\bibinfo {year} {2011})}\BibitemShut {NoStop}%
\bibitem [{\citenamefont {Pechenick}\ \emph {et~al.}(2015)\citenamefont
  {Pechenick}, \citenamefont {Danforth},\ and\ \citenamefont
  {Dodds}}]{Pechenick2015}%
  \BibitemOpen
  \bibfield  {author} {\bibinfo {author} {\bibfnamefont {E.~A.}\ \bibnamefont
  {Pechenick}}, \bibinfo {author} {\bibfnamefont {C.~M.}\ \bibnamefont
  {Danforth}},\ and\ \bibinfo {author} {\bibfnamefont {P.~S.}\ \bibnamefont
  {Dodds}},\ }\href {https://doi.org/10.1371/journal.pone.0137041} {\bibfield
  {journal} {\bibinfo  {journal} {PLoS ONE}\ }\textbf {\bibinfo {volume}
  {10}},\ \bibinfo {pages} {1} (\bibinfo {year} {2015})},\ \Eprint
  {https://arxiv.org/abs/1501.00960} {arXiv:1501.00960} \BibitemShut {NoStop}%
\bibitem [{\citenamefont {Michel}\ \emph {et~al.}(2011)\citenamefont {Michel},
  \citenamefont {Shen}, \citenamefont {Aiden}, \citenamefont {Veres},
  \citenamefont {Gray}, \citenamefont {Pickett}, \citenamefont {Hoiberg},
  \citenamefont {Clancy}, \citenamefont {Norvig}, \citenamefont {Orwant},
  \citenamefont {Pinker}, \citenamefont {Nowak},\ and\ \citenamefont
  {Aiden}}]{Michel2011}%
  \BibitemOpen
  \bibfield  {author} {\bibinfo {author} {\bibfnamefont {J.-B.}\ \bibnamefont
  {Michel}}, \bibinfo {author} {\bibfnamefont {Y.~K.}\ \bibnamefont {Shen}},
  \bibinfo {author} {\bibfnamefont {A.~P.}\ \bibnamefont {Aiden}}, \bibinfo
  {author} {\bibfnamefont {A.}~\bibnamefont {Veres}}, \bibinfo {author}
  {\bibfnamefont {M.~K.}\ \bibnamefont {Gray}}, \bibinfo {author}
  {\bibfnamefont {J.~P.}\ \bibnamefont {Pickett}}, \bibinfo {author}
  {\bibfnamefont {D.}~\bibnamefont {Hoiberg}}, \bibinfo {author} {\bibfnamefont
  {D.}~\bibnamefont {Clancy}}, \bibinfo {author} {\bibfnamefont
  {P.}~\bibnamefont {Norvig}}, \bibinfo {author} {\bibfnamefont
  {J.}~\bibnamefont {Orwant}}, \bibinfo {author} {\bibfnamefont
  {S.}~\bibnamefont {Pinker}}, \bibinfo {author} {\bibfnamefont {M.~a.}\
  \bibnamefont {Nowak}},\ and\ \bibinfo {author} {\bibfnamefont {E.~L.}\
  \bibnamefont {Aiden}},\ }\href {https://doi.org/10.1126/science.1199644}
  {\bibfield  {journal} {\bibinfo  {journal} {Science (New York, N.Y.)}\
  }\textbf {\bibinfo {volume} {331}},\ \bibinfo {pages} {176} (\bibinfo {year}
  {2011})}\BibitemShut {NoStop}%
\bibitem [{\citenamefont {Zhigljavsky}(2011)}]{Zhigljavsky2011}%
  \BibitemOpen
  \bibfield  {author} {\bibinfo {author} {\bibfnamefont {A.}~\bibnamefont
  {Zhigljavsky}},\ }\href {https://doi.org/10.1007/978-3-642-04898-2_521}
  {\emph {\bibinfo {title} {International Encyclopedia of Statistical
  Science}}}\ (\bibinfo  {publisher} {Springer},\ \bibinfo {year} {2011})\ pp.\
  \bibinfo {pages} {1335--1337}\BibitemShut {NoStop}%
\bibitem [{\citenamefont {Alexandrov}\ and\ \citenamefont
  {Golyandina}(2005)}]{alexandrov2005automatic}%
  \BibitemOpen
  \bibfield  {author} {\bibinfo {author} {\bibfnamefont {T.}~\bibnamefont
  {Alexandrov}}\ and\ \bibinfo {author} {\bibfnamefont {N.}~\bibnamefont
  {Golyandina}},\ }in\ \href@noop {} {\emph {\bibinfo {booktitle} {Proceedings
  of the 5th St. Petersburg Workshop on Simulation}}}\ (\bibinfo {organization}
  {St. Petersburg State University St. Petersburg},\ \bibinfo {year} {2005})\
  pp.\ \bibinfo {pages} {45--50}\BibitemShut {NoStop}%
\bibitem [{\citenamefont {Karjus}\ \emph {et~al.}(2020)\citenamefont {Karjus},
  \citenamefont {Blythe},\ and\ \citenamefont {Kirby}}]{Karjus2020}%
  \BibitemOpen
  \bibfield  {author} {\bibinfo {author} {\bibfnamefont {A.}~\bibnamefont
  {Karjus}}, \bibinfo {author} {\bibfnamefont {R.~A.}\ \bibnamefont {Blythe}},\
  and\ \bibinfo {author} {\bibfnamefont {S.}~\bibnamefont {Kirby}},\ }\href
  {https://doi.org/10.1163/22105832-01001200} {\bibfield  {journal} {\bibinfo
  {journal} {Language Dynamics and Change}\ }\textbf {\bibinfo {volume} {10}},\
  \bibinfo {pages} {86} (\bibinfo {year} {2020})},\ \Eprint
  {https://arxiv.org/abs/1806.00699} {arXiv:1806.00699} \BibitemShut {NoStop}%
\bibitem [{\citenamefont {Matsuda}\ and\ \citenamefont
  {Komaki}(2017)}]{Matsuda2017}%
  \BibitemOpen
  \bibfield  {author} {\bibinfo {author} {\bibfnamefont {T.}~\bibnamefont
  {Matsuda}}\ and\ \bibinfo {author} {\bibfnamefont {F.}~\bibnamefont
  {Komaki}},\ }\href {https://doi.org/10.1162/NECO} {\bibfield  {journal}
  {\bibinfo  {journal} {Neural Computation}\ }\textbf {\bibinfo {volume}
  {29}},\ \bibinfo {pages} {1} (\bibinfo {year} {2017})},\ \Eprint
  {https://arxiv.org/abs/1803.01446} {arXiv:1803.01446} \BibitemShut {NoStop}%
\bibitem [{Note2()}]{Note2}%
  \BibitemOpen
  \bibinfo {note} {\protect \href
  {http://storage.googleapis.com/books/ngrams/books/datasetsv2.html}{http://storage.googleapis.com/books/ngrams/books/}
  datasetsv2.html}\BibitemShut {NoStop}%
\bibitem [{Note3()}]{Note3}%
  \BibitemOpen
  \bibinfo {note} {\protect \href
  {https://github.com/AlePardoPintos/Words-Phase-Coupled}{https://github.com/AlePardoPintos/Words-Phase-Coupled}}\BibitemShut
  {NoStop}%
\end{thebibliography}%

\end{document}